\documentstyle[preprint,tighten,aps]{revtex}
\begin{document}
\draft
\preprint{MIT-CTP\#2301}
\title{Statistical mechanics of \\
  relativistic anyon-like systems \footnotemark[1]}

\footnotetext[1]{This work is supported in part by funds
provided by the U. S. Department of Energy (D.O.E.) under
contracts \#DE-FG06-88ER40427, \#DE-AC02-76ER03069 and by a
Feodor-Lynen-Fellowship of the Alexander von Humboldt-Foundation
(B.S.).}

\author{Suzhou~Huang}
\address{Department of Physics, FM--15, University of Washington\\
         Seattle, Washington ~98195\\
         {\rm and}
         Center for Theoretical Physics,
         Laboratory for Nuclear Science
         and Department of Physics\\
         Massachusetts Institute of Technology,
         Cambridge, Massachusetts 02139}
\author{Bernd~Schreiber}
\address{Center for Theoretical Physics,
         Laboratory for Nuclear Science
         and Department of Physics\\
         Massachusetts Institute of Technology,
         Cambridge, Massachusetts 02139}

\date{\today}
\maketitle

\begin{abstract}

  To study the manifestation of the Aharonov-Bohm effect in
many-body systems we consider the statistical mechanics of the
Gross-Neveu model on a ring (1+1 dimensions) and on a cylinder
(2+1 dimensions) with a thin solenoid coinciding with the
axis. For such systems with a non-trivial magnetic flux
($\theta$) many thermodynamical observables, such as the order
parameter, induced current and virial coefficients, display
periodic but non-analytic dependence on $\theta$. In the 2+1
dimensional case we further find that there is an interval of
$\theta\in(1/3,2/3)$ (modulo integers) where parity is always
spontaneously broken, independent of the circumference of the
cylinder. We show that the mean-field character of the phase
transitions is preserved to the leading order in $1/N$, by
verifying the $\theta$-independence of all the critical
exponents. The precise nature of the quasi-particle, locally
fermion-like and globally anyon-like, is illuminated through the
calculation of the equal-time commutator and the decomposition
of the propagator into a sum  over paths classified by winding
numbers.

\end{abstract}

\widetext

\section{Introduction}

  Due to their direct relevance to the quantum Hall effect
\cite{shapere}, possible connection with high temperature
superconductivity \cite{wilczek} as well as their intellectual
challenges, anyon systems have attracted enormous attention
in the past several years. Almost exclusively, the formulation
of anyons is based on the idea of the Chern-Simons
construction. In this formulation anyons are locally
so-called charge-flux tube composites. Physically, the
essence involved is nothing but the famous Aharonov-Bohm
effect \cite{aharonov}.
It is the interference of the gauge field, (which would
be a pure gauge when totally isolated), associated with each
particle that modifies the statistics of the particle
such that it interpolates
between the boson and fermion limits. While it is conceptually more
satisfactory, the involvement of the Chern-Simons term, because
of its technical complexity, makes the understanding of anyon
properties rather difficult, especially in many-anyon systems.

  In this paper we study a system which in many ways captures
the main features of anyons, while avoiding the inclusion of the
Chern-Simons term. In a topologically simple configuration
space it would be necessary to include the Chern-Simons term
to implement the fractional statistics. However, when the
configuration space becomes topologically non-trivial, there
is no need to add terms other than the minimal coupling between
the gauge and the matter fields in order to achieve the desired
interference. Thus with the  slight complication of introducing a
topologically non-trivial configuration space, the procedure
of setting up a system which  displays anyon-like behavior
is greatly simplified and one can study typical effects
such as the periodic but non-analytic
dependence of thermodynamical observables on the
statistics parameter. It is this technical simplification which
allowed us to carry out our study of the thermodynamics of such
a system rather explicitly, using conventional methods. We hope
that our results will shed some light on the thermodynamics of
Chern-Simons anyon systems.

The specific model we consider is the Gross-Neveu model
\cite{gross} with an external constant gauge potential defined by
\begin{equation}
{\cal L}=\bar{\psi}[i\partial_\mu-a_\mu(\theta)]\gamma^\mu\psi
+{g^2\over 2N}(\bar{\psi}\psi)^2 \, ,
\label{model}
\end{equation}
on a ring (1+1 dimensions) and on a cylinder (2+1 dimensions)
with $x$-direction compactified (of length $L$) to the leading
order in $1/N$. The $\psi$-field is a two-component Dirac
spinor and implicitly has $N$ components in the internal space.
The external gauge field $a_\mu(\theta)$ is
generated by a thin solenoid of magnetic flux $2\pi\theta$
which coincides with the axis of the ring or cylinder, that is,
$a_1(\theta)=2\pi\theta/L$ and all other components vanish.
Since the Lagrangian is fermionic when $\theta=0$, the
anti-periodic boundary condition in $x$-direction,
$\psi(x+L)=-\psi(x)$, is naturally imposed. Without losing
generality we only need to consider $\theta\in [0,1)$, for the
integral part of $\theta$ can be safely gauge-transformed away.
The periodic dependence of physical quantities on $\theta$ is
obvious.

  The geometric setting here is such that the particles move in
a multi-connected configuration space. Due to the presence of
the magnetic flux, we anticipate that the Aharonov-Bohm effect
will have a profound impact on the statistical mechanics of the
system. It is possible to formally remove the gauge field in
Eq.(\ref{model}) via a gauge transformation
$\psi\rightarrow \exp(i2\pi\theta x/L)\psi\equiv\psi'$. However,
the expense of eliminating the gauge field is that the new field
$\psi'$ becomes multi-valued (when $\theta\neq 0$ or $1/2$).
The magnetic flux can be regarded as an external means of varying
the boundary condition. It should be emphasized that, like in the
original Aharonov-Bohm experiment, the
magnetic field on the ring or cylinder is zero. Therefore, the
effect of $\theta\neq 0$ is purely quantum mechanical in nature,
taking the advantage of $L$ being finite. In the limit
$L\rightarrow\infty$ all the $\theta$-dependence of physical
observables should disappear,
since then the effect of the boundary condition becomes irrelevant.
It is also easy to see that single particle states have smooth
$\theta$-dependence and that the non-analyticity of physical observables
arises only from many-body effects.

  Before we dwell into details, let us comment on the choice
of the specific microscopic Lagrangian, Eq.(\ref{model}). In our
opinion, the precise underlying dynamics is not too crucial, as
long as the qualitative features, such as the existence of a phase
transition, are present. Therefore, the choice is simply dictated
by convenience. However, it should be pointed out that the
Gross-Neveu type of Lagrangians do have some relevance to certain
condensed matter systems or models. For example, the Gross-Neveu
model in 1+1 dimensions is closely related to the continuum limit
of the one dimensional Hubbard model at half-filling and the quantum
spin-one-half antiferromagnetic Heisenberg chain \cite{hubbard},
while in 2+1 dimensions it models the continuum limit of the
so-called chiral spin liquid \cite{wen}. Although we do not directly
address questions on how to make observations in some specifically
designed experiments, one should always keep these issues in mind.

  The Lagrangian in Eq.(\ref{model}) possesses discrete symmetries.
In the 1+1 dimensional case, the symmetry is the chiral symmetry
\begin{equation}
\psi(x,t)\rightarrow e^{i\gamma_5\pi/2}\psi(x,t) \, ,
\label{chiral}
\end{equation}
whereas in the 2+1 dimensional case, the relevant symmetry is  parity
\begin{equation}
\psi(x,y,t)\rightarrow i\gamma_2\psi(x,-y,t) \, .
\label{parity}
\end{equation}
Since the operator $\bar{\psi}\psi$ changes sign under either
Eq.(\ref{chiral}) or Eq.(\ref{parity}), the order parameters are the
ensemble averages $\langle\bar{\psi}\psi\rangle$. In other words,
when a mass term is generated, the discrete symmetry, the chiral
symmetry in 1+1 dimensions or parity in 2+1 dimensions, is
dynamically broken. Because of the discrete nature of the symmetry
involved, there will be no Goldstone mode associated with the
dynamical symmetry breaking in our model. For convenience, we
introduce a scalar auxiliary field $\sigma$, which interpolates
the composite operator $-g^2\bar{\psi}\psi$. Then the Lagrangian
in Eq.(\ref{model}) is equivalent to
\begin{equation}
{\cal L}=\bar{\psi}[i\partial_\mu-a_\mu(\theta)]\gamma^\mu\psi
-{N\over 2g^2}\sigma^2-\sigma\bar{\psi}\psi \, .
\label{model1}
\end{equation}
Note that the $\sigma$-field obeys periodic boundary conditions for
any value of $\theta$, though it is odd under the transformations
Eq.(\ref{chiral}) or Eq.(\ref{parity}).

  The rest of the paper is organized as follows. In the next section
we present the result in 1+1 dimensions, along with its relevance
to L\"{u}scher's small-volume expansion in asymptotically free
field theories with phase transitions. In section III we present
the result in 2+1 dimensions, in a manner suited for possible
verification in condensed matter experiments. To further reveal
the effect of $\theta$ on the phase transition, we compute in
section IV all the critical exponents and verify that the phase
transitions remain in the universality class of mean-field type to
leading order in $1/N$. The anyonic character of the quasi-particle
is demonstrated in section V via an explicit calculation of the
relevant propagators and commutators. Finally, we summarize and
point out possible generalizations in section VI.

\section{Results in 1+1 dimensions}

  To leading order in $1/N$ the effective potential for
the $\sigma$-field is given by the standard one-loop formula, in
dimensional regularization,
\begin{equation}
V_{\rm eff}[\sigma,\theta]=N\biggl[
{\mu^{2\epsilon}\sigma^2\over 2g_B^2}-
{1\over L}\sum_{n=-\infty}^\infty \mu^{2\epsilon}\int
{d^{1-2\epsilon}\omega \over (2\pi)^{1-2\epsilon}}
\ln(\omega^2+\tilde{k}_n^2+\sigma^2)\biggr] \, ,
\label{veff2}
\end{equation}
where $\tilde{k}_n=k_n+2\pi\theta/L$ and $k_n=\pi(2n+1)/L$. Since
an additive constant in $V_{\rm eff}[\sigma,\theta]$ is irrelevant,
we can first calculate
\begin{equation}
{\partial V_{\rm eff}[\sigma,\theta]\over\partial\sigma}
=N\sigma\biggl[{\mu^{2\epsilon}\over g_B^2}
-{1\over L}\sum_{n=-\infty}^\infty \mu^{2\epsilon}\int
{d^{1-2\epsilon}\omega \over (2\pi)^{1-2\epsilon}}
{1\over\omega^2+\tilde{k}_n^2+\sigma^2} \biggr] \, .
\end{equation}
After some straightforward manipulation we obtain explicitly
\begin{equation}
{\partial V_{\rm eff}[\sigma,\theta]\over\partial\sigma}
={N\sigma\over\pi}\biggl[{\pi\mu^{2\epsilon}\over g_B^2}
-{1\over 2}\Bigl({1\over\epsilon}+\gamma_E-\ln{4\pi}
+\ln(\mu^2L^2)\Bigr)-C_0(\theta)-\sum_{\nu=1}^\infty
C_{2\nu}(\theta) (\sigma L)^{2\nu}\biggr] \, ,
\end{equation}
where
\begin{equation}
C_0(\theta)={1\over 2}\sum_{\nu=1}^\infty \zeta(1+\nu)
\Bigl[ (1/2+\theta)^\nu+(1/2-\theta)^\nu\Bigr]
={1\over 2}\sum_{n=1}^\infty\Bigl[
{1/2-\theta\over n(n-1/2+\theta)}
+{1/2+\theta\over n(n-1/2-\theta)}\Bigr]\, ,
\label{c0}
\end{equation}
and
\begin{equation}
C_{2\nu}(\theta)={(-1)^\nu\over 2}
\left(\begin{array}{c} 2\nu \\ \nu \end{array}\right)
\Bigl({1\over 4\pi}\Bigr)^{2\nu}
\Bigl[\zeta(2\nu+1,1/2+\theta)+\zeta(2\nu+1,1/2-\theta)\Bigr]\, .
\end{equation}
In the above equation $\gamma_E$ is the Euler-Mascheroni constant
and $\zeta(\alpha,z)=\sum_{n=0}^\infty (n+z)^{-\alpha}$ is the
generalized Riemann Zeta function.
By further choosing the modified minimal subtraction scheme,
the final result for the renormalized effective potential is
then given by
\begin{equation}
{\partial V_{\rm eff}[\sigma,\theta]\over\partial\sigma}
={N\sigma\over\pi}\biggl[-\ln(ML)-\gamma_E+\ln{4\pi}
-C_0(\theta)-\sum_{\nu=1}^\infty
C_{2\nu}(\theta) (\sigma L)^{2\nu}\biggr] \, ,
\label{dveff2}
\end{equation}
where $M$ is a shorthand notation for the standard
$\Lambda_{\overline{\rm MS}}$. Since the convergence radius of
the series in $\sigma L$ in the above equation is $2\pi$, it is
sometimes necessary to use the integral representation of
the series
\begin{equation}
C_0(\theta)+\sum_{\nu=1}^\infty C_{2\nu}(\theta) z^{2\nu}=
\ln{4\pi}-\gamma_E-\ln{z}+\sum_{n=1}^\infty (-1)^n \cos(2\pi n\theta)
\int_0^\infty{dt\over t}\exp\bigl(-z^2t-{n^2\over 4t}\bigr)\, ,
\end{equation}
in Eq.(\ref{dveff2}).

  The mass gap is solved from
$\partial V_{\rm eff}(\sigma=m)/\partial\sigma=0$, yielding
\begin{equation}
m(L,\theta)=\left\{\begin{array}{ll}
M\exp\Bigl[\sum_{n=1}^\infty (-1)^n \cos(2\pi n\theta)
\int_0^\infty {dt\over t} \exp(-z^2 t-n^2/4t)\Bigr] \, ,
&\text{when $L>L_c(\theta)$,}\\
0\, , &\text{when $L<L_c(\theta)$,}\end{array} \right.
\label{gap2}
\end{equation}
where $z\equiv m(L,\theta)L$ and $L_c(\theta)$ is the critical
length of the ring defined by
\begin{equation}
L_c(\theta)={\pi\over M}
\exp\Bigl[-\gamma_E+\ln{4}-C_0(\theta)\Bigr] \, .
\label{lc2}
\end{equation}
As defined in Eq.(\ref{c0}) $C_0(\theta)$ is a monotonically
increasing function of $\theta\in[0,1/2)$ starting at
$C_0(0)=\ln{4}$ and diverging  at
$\theta=1/2$. Thus the longest critical length results  when
$\theta=0$, $L_c(0)=\pi\exp(-\gamma_E)/M\approx 1.7639/M$, and
the shortest critical length is zero, $L_c(1/2)=0$. For other
values of $\theta$ the critical length interpolates
continuously between these two extrema. In Fig.\ref{fig1} we
plot $m(L,\theta)$ as a function of $1/L$ for several values
of $\theta$.

  The non-trivial dependence of the mass gap on $\theta$ and
$L$ has an interesting implication with respect to the small-volume expansion,
proposed by L\"{u}scher \cite{luescher,luescher1},
for asymptotically free theories with
phase transitions. L\"{u}scher's approach is based
on the observation that when the volume is small, the running
coupling becomes small due to
asymptotic freedom, and perturbative calculations can then be
applied. The physical result is finally obtained by doing a
sequence of calculations with increasing volumes and extrapolating
to the infinite volume limit. However, there are some obstacles
in doing the infinite volume extrapolation. One is due to the
existence of instantons, mostly addressed so far in the literature
\cite{baal}. The other is related to the existence of phase
transitions, such as the deconfinement transition in QCD at
finite temperature. When a phase transition is present
the infinite volume extrapolation can hardly be smooth. The 1+1
dimensional Gross-Neveu model in the large-$N$ limit provides
a perfect example for illustrating the problem.

  Since $M=\Lambda_{\overline{\rm MS}}$ is the only dimensionful
parameter, the transition point $L_c$ is of order one in
units of $1/M$ when $\theta=0$. This transition is of course
the same as the finite temperature phase transition. From
Fig.\ref{fig1} we see that the entire regime of finite mass
gap is located in the non-perturbative region, where the perturbative
running coupling is very large or divergent, as exemplified by
the first order (dotted line) and second order (dashed line)
perturbative results, respectively. When the size of the box $L$
is smaller than $L_c$, chiral symmetry forbids the mass gap
generation. Hence the small-volume expansion in the most naive
form does not work at all. However, as we have shown, we can
delay the phase transition to an arbitrary point by varying the
relevant boundary condition. This widens the window in which  the
mass gap curve overlapps with the perturbative regime. More
concretely, let us examine the mass gap behavior near the bosonic
limit. To order $(\theta-1/2)^2$ the mass gap equation is
\begin{equation}
\ln\Bigl({1\over ML}\Bigr)+\ln{4\pi}-\gamma_E=
\Bigl(4(\theta-1/2)^2+{m(L,\theta)^2L^2\over\pi^2}\Bigr)^{-1/2}\, ,
\end{equation}
which yields, in the region of $|\theta-1/2|<g^2(L)/2\pi$,
\begin{equation}
m(L,\theta)={g^2(L)\over L}
\sqrt{1-{4\pi^2(\theta-1/2)^2\over g^2(L)}}\, ,
\end{equation}
where $g^2(L)=\pi/\ln(4\pi e^{-\gamma_E}/ML)$ is the perturbative
running coupling constant. The critical region is then shifted to
$g^2(L)\sim|\theta-1/2|$. Thus, close to the bosonic
limit $\theta=1/2$, the mass gap curve possesses  a big region
in which a  perturbative calculation can be carried out. At the
very value of $\theta=1/2$ the gap equation Eq.(\ref{dveff2})
becomes identical to that of the non-linear $\sigma$-model in
the large-$N$ limit \cite{luescher}. In practice,
some intermediate values of $\theta$ may be optimal to make the
curve flat enough for a better extrapolation, though the choice of
$\theta$ would only be known {\it a posteriori}. Using more
general boundary conditions, such as twisted boundary conditions
\cite{thooft}, to improve the convergence in the Yang-Mills gauge
theory have been pursued \cite{twisted}.

  Strictly speaking, the analogy with finite temperature phase
transitions is no longer valid when more than one spatial
directions are compactified. However, even in these cases, rapid
cross-overs are still expected, though there would be no true
singularities, which only associate with phase transitions.

  We would like to mention in passing that the phase transition
in the 1+1 dimensional Gross-Neveu model with a finite $L$ is no
longer possible when $N$ is finite due to the existence of the
$\sigma$-kinks, as shown in Ref. \cite{dashen}.
For this reason we will not dwell further into details on
the thermodynamical observables in the 1+1 dimensional case.
However, since this absence of the phase transition is invisible
in the large-$N$ expansion to any finite order, the results we
found above are entirely legitimate within the domain of the $1/N$
expansion. As we will show in the next section, qualitatively
similar results are obtained in the 2+1 dimensional Gross-Neveu
model, where a phase transition can exist for an infinitely  long
cylinder (but with finite circumference $L$) at zero temperature.

\section{Results in 2+1 dimensions}

The effective potential for the $\sigma$-field in 2+1 dimensions
to leading order in $1/N$ has a similar form as Eq.(\ref{veff2})
\begin{equation}
V_{\rm eff}[\sigma,\theta]=N\biggl[
{\mu^{2\epsilon}\sigma^2\over 2g_B^2}-
{1\over L}\sum_{n=-\infty}^\infty \mu^{2\epsilon}\int
{d^{2-2\epsilon}\omega \over (2\pi)^{2-2\epsilon}}
\ln(\omega^2+\tilde{k}_n^2+\sigma^2)\biggr] \, .
\label{veff3}
\end{equation}
Again we first calculate the derivative of $V_{\rm eff}$
with respect to $\sigma$. After some standard manipulation
we obtain
\begin{equation}
{\partial V_{\rm eff}[\sigma,\theta]\over\partial\sigma}
=N\sigma\biggl[{\mu^{2\epsilon}\over g_B^2}
+{1\over 2\pi L}\ln(4\cos^2{\pi\theta})+{1\over 2\pi L}
\sum_{\nu=1}^\infty C_{2\nu}(\theta) (\sigma L)^{2\nu}\biggr]\, ,
\label{dveff3}
\end{equation}
where
\begin{equation}
C_{2\nu}(\theta)={(-1)^{\nu+1}\over\nu}
\Bigl({1 \over 2\pi}\Bigr)^{2\nu}
\Bigl[\zeta(2\nu,1/2+\theta)+\zeta(2\nu,1/2-\theta)\Bigr]\, .
\end{equation}
Due to the use of dimensional regularization we do not encounter
infinities here. However, in order for the theory to possess  spontaneous
parity breaking in the limit $L\rightarrow\infty$ we demand, as
in Ref. \cite{review},
\begin{equation}
{\mu^{2\epsilon}\over g_B^2}=-{M\over 2\pi} \, .
\end{equation}
$M$ will turn out to be the dynamical mass of the particle in the
limit $L\rightarrow\infty$. We want to remind the reader that in
2+1 dimensions $g_B^2$ has the  dimension of inverse mass. If we further
use the infinite product representation of
\begin{equation}
\cosh{z}-\cos{2\pi\alpha}=
2\sin^2{\pi\alpha}\Bigl(1+{z^2\over 4\pi^2\alpha^2}\Bigr)
\prod_{k=1}^\infty\Bigl(1+{z^2\over 4\pi^2(k+\alpha)^2}\Bigr)
\Bigl(1+{z^2\over 4\pi^2(k-\alpha)^2}\Bigr) \, ,
\end{equation}
the series in Eq.(\ref{dveff3}) can be summed into a closed form
\begin{equation}
{\partial V_{\rm eff}[\sigma,\theta]\over\partial\sigma}
={N\sigma\over 2\pi}\biggl[-M+{1\over L}
\ln\bigl(2\cosh{\sigma L}+2\cos{2\pi\theta}\bigr)\biggr]\, .
\end{equation}
The mass gap is immediately solved from
$\partial V_{\rm eff}(\sigma=m)/\partial\sigma=0$,
\begin{equation}
m(L,\theta)=\left\{\begin{array}{ll}
{1\over L}\cosh^{-1}
\Bigl[1+{1\over 2}(e^{ML}-e^{ML_c(\theta)})\Bigr]\, ,
&\text{when $L>L_c(\theta)$,}\\
0\, , &\text{when $L<L_c(\theta)$,}\end{array} \right.
\label{gap3}
\end{equation}
where the critical length $L_c(\theta)$ is defined by
\begin{equation}
L_c(\theta)=\left\{\begin{array}{ll}
0\, , &\text{when $\theta\in(1/3,2/3)$,} \\
{1\over M}\ln(4\cos^2{\pi\theta})\, ,
&\text{otherwise.}\end{array}\right.
\label{lc3}
\end{equation}
The free energy, equal to the effective potential evaluated
at $\sigma=m(L,\theta)$, has the form
\begin{equation}
F[L,\theta]={N\over 2\pi}\biggl[-{M\over 2}m^2(L,\theta)
+{1\over L}\int_0^{m(L,\theta)}d\sigma\,\sigma
\ln\bigl(2\cosh{\sigma L}+2\cos{2\pi\theta}\bigr)\biggr]\, .
\label{fe3}
\end{equation}

  For convenience from the experimental point of view we would like
to regard $\theta$, rather than $1/L$, as the independent variable,
because one could use the same sample and only change the magnetic
field strength in the solenoid. In Fig.\ref{fig2} we display the
mass gap as a function of $\theta$ for three values of $1/L$. The
distinct feature in the figure is the non-analyticity of
$m(L,\theta)$ as a function of $\theta$ when $L\le L_c(0)$, reflecting
the fact that the Aharonov-Bohm effect pushes the critical length
of the phase transition to a smaller value or effectively changes
the mass scale of the theory to a larger value. In Fig.\ref{fig3}
we plot the phase diagram in the $(1/L,\theta)$-plane determined from
Eq.(\ref{gap3}). We not only find the change of the critical length
as $\theta$ is varied, we also find, surprisingly, that there is
a region $\theta\in(1/3,2/3)$ where parity is always broken,
no matter how small $L$ is. In other words, even when the
renormalized mass scale $M=0$, the system would still prefer
to stay in the parity breaking phase for $\theta\in(1/3,2/3)$.

  To understand why there should be such a region of $\theta$
where parity is always broken, let us focus on the
$\sigma$-propagator. We know from the literature \cite{review}
that the mechanism of parity breaking at zero temperature
(corresponding to $L\rightarrow\infty$ and $\theta=0$ in our
case) is due to the formation of a quasi-bound state, which
shows up as a ``pole'' at the fermion-antifermion threshold,
$\omega^2=4M^2$. It is this quasi-bound state,
analogous to the Cooper-pair in an usual superconductor,
that condenses and renders the expectation value
$\langle\bar{\psi}\psi\rangle$ non-vanishing. By an explicit
calculation of the $\sigma$-propagator, we can show that this
quasi-bound state persists whenever the mass gap $m(L,\theta)$ is
finite. Therefore the mechanism for  parity breaking is
the same when $\theta\in(1/3,2/3)$ as when $L\rightarrow\infty$
and $\theta=0$.
Since the threshold becomes zero in the symmetric phase
due to the vanishing mass gap, all states have to lie
within the continuum and hence there can not be any bound state.
One can easily verify that $\omega^2=0$ is no longer a pole for
the $\sigma$-propagator in the symmetric phase.

  Qualitatively, the symmetric phase is a kinetic energy
dominated phase, while the symmetry breaking phase is the
interaction dominated phase. When the size of the cylinder $L$
becomes smaller, the kinetic energy becomes more and more important,
if none of the discrete momentum levels vanish. In our case the
effective momentum is $\tilde{k}_n=\pi(2n+1+\theta)/L$, which
interpolates between the fermionic and bosonic cases. When $\theta$ is
finite, one of the $\tilde{k}_n$ is closer to zero and
thus the role of the kinetic energy is reduced. This in turn implies
that, effectively, the interaction becomes  stronger when $\theta$
increases from the fermionic value 0 to the bosonic value 1/2.
We will show, in section V, that the presence of $\theta$ indeed
modifies the phase acquired when interchanging two particles,
from the original fermionic value, $\pi$, to $\pi+2\pi\theta$. This
extra phase could be interpreted as an additional contact interaction.
One such example in the context of non-relativistic quantum mechanics
was noticed some time ago by Leinaas and Myrheim \cite{leinaas}.

  Another physical observable is the induced (or persistent)
surface current density circling around the cylinder, defined by
\begin{equation}
J[L,\theta]\equiv -{\partial F[L,\theta]\over\partial a_1}=
-{L\over 2\pi}{\partial F[L,\theta]\over \partial\theta}
={N\sin{2\pi\theta}\over 2\pi L^2}\int_0^{m(L,\theta)L} dx
{x\over\cosh{x}+\cos{2\pi\theta}} \, .
\end{equation}
One can easily recognize that the overall factor $\sin{2\pi\theta}$
in the above equation has the same origin as a similar
factor in the Aharonov-Bohm scattering amplitude \cite{aharonov}.
This overall factor implies that the induced current vanishes at
the fermion and boson limits or when $\theta$ is integer or half
integer, in analogy  to the vanishing of the Aharonov-Bohm scattering
cross section in the same limits. A similar induced current on a
ring was derived in terms of a single particle picture
in Ref. \cite{gerry} more than a decade ago.
What is novel in our case is that $m(L,\theta)$ is a dynamically
determined quantity, due to many-body effects, rather than an
external parameter. In Fig.\ref{fig4} we plot $J[L,\theta]$ as a
function of $\theta$ for several values of $1/L$. Again we observe
the non-analyticity, inherent from the non-analyticity of
$m(L,\theta)$.

  Expanding the free energy in Eq.(\ref{fe3}) as a power series
in  $m(L,\theta)L$, we obtain
\begin{equation}
F[L,\theta]={N m(L,\theta)^2\over 4\pi}
\biggl[-M+{1\over L}\ln(4\cos^2{\pi\theta})+{1\over L}
\sum_{\nu=1}^\infty \overline{C}_{2\nu}(\theta)
\Bigl(m(L,\theta)L\Bigr)^{2\nu}\biggr] \, ,
\end{equation}
with
\begin{equation}
\overline{C}_{2\nu}(\theta)=
{(-1)^{\nu+1}\over \nu(\nu+1)}\Bigl({1 \over 2\pi}\Bigr)^{2\nu}
\Bigl[\zeta(2\nu,1/2+\theta)+\zeta(2\nu,1/2-\theta)\Bigr]\, .
\end{equation}
This expression for $F[L,\theta]$ can be interpreted as a virial
expansion, in analogy to the expansion of pressure as a power series
in the particle density. The corresponding virial coefficients
are $\overline{C}_{2\nu}(\theta)$, which become singular when
$\theta=1/2$, since $\zeta(\alpha,z)\sim z^{-\alpha}$ when $z$
is small enough.

  We will show in section V that $\theta$ plays the role of the
statistics parameter, with $\theta=0$ being the fermionic limit
and $\theta=1/2$ being the bosonic limit. The divergence of the
virial coefficients at $\theta=1/2$ apparently reflects the fact
that it is not possible to use a single series to represent the
free energy with all the coefficients smoothly interpolating
between the fermionic and bosonic regimes, even though both regimes
are well behaved by themselves. Put differently, the fermionic
regime and bosonic regime are separated by non-perturbative physics
(in terms of the statistics parameter). A much milder singularity
(cusp) is found for the second virial coefficient in the
non-relativistic anyon system considered by Arovas, Schrieffer,
Wilczek and Zee \cite{arovas}. One may  speculate that
the stronger singularity we find is rooted in the deep connection
between statistics and spin in fully relativistic theories. It is
worth emphasizing again that the non-analyticity is a manifestation
of the Aharonov-Bohm effect in an infinite many-particle system.
Without a phase transition, the $\theta$ dependence of physical
observables would never be singular.

\section{Universality}

  In the last two sections we found that many thermodynamical
observables are strongly influenced by the Aharonov-Bohm effect.
In order to reveal the precise influence  of the Aharonov-Bohm
effect on the phase transition we calculate the critical exponents
in this section. It turns out that if we fix $\theta$
and vary $1/L$, in analogy  to the usual variation of temperature,
the universalities of the chiral restoration in the 1+1 dimensional
case and parity restoration in the 2+1 dimensional case are not
altered by the presence of $\theta$ to leading order in $1/N$, as
long as the phase transition is not totally destroyed. Since the
calculation process is insensitive to the space-time dimensions
involved, we will only present the calculation in the case of 2+1
dimensions. For notational simplicity let us define
$\tau\equiv [L-L_c(\theta)]/L_c(\theta)$, with $L_c(\theta)$ given
by Eq.(\ref{lc3}). For detailed definitions
of the critical exponents we refer to the book \cite{itzykson}.
Because the phase transition exists when $\theta\in [0,1/3)$
and $\theta\in (2/3,1]$, the critical exponents are defined only
in these regions in 2+1 dimensions.

  From Eq.(\ref{gap3}) we can immediately solve for the mass gap
$m(L,\theta)$, for small $\tau>0$,
\begin{equation}
m(L,\theta)\approx
{M\over\sqrt{\ln(4\cos^2{\pi\theta})} } \, \tau^{1/2},
\end{equation}
which in turn yields $\beta=1/2$.

  The free energy clearly vanishes in the symmetric phase when
$L<L_c(\theta)$, since $m(L,\theta)=0$ there.
In the symmetry breaking phase the free energy is given, as can be seen
directly from Eq.(\ref{fe3}), by
\begin{equation}
F[L,\theta]\approx{N m(L,\theta)^2\over 4\pi}
\biggl[-M+{L_c(\theta)\over L}M+
\overline{C}_2(\theta)\Bigl(m(L,\theta)L\Bigr)^2\biggr]
=-{N M^3\over 4\pi}
{1-\overline{C}_2(\theta)\over\ln(4\cos^2{\pi\theta})}\, \tau^2\, ,
\end{equation}
with $\overline{C}_2(\theta)=1/8\cos^2\pi\theta$.
The specific heat $c_v$ is proportional to the second derivative
of $F[L,\theta]$ with respect to $\tau$. Thus we see that $c_v$ is
finite in both phases and has a discontinuity at the critical point,
which implies that $\alpha=\alpha'=0$.

  To obtain the exponents $\gamma$, $\nu$ and $\eta$ we need to
calculate the two-point correlation function for the $\sigma$-field
in momentum space in the ``static'' limit,
$D_\sigma(\omega=0,{\bf p})$. Here $\omega$ should be understood as
the momentum in the  $x$-direction and ${\bf p}$ represents the momenta along
the Euclidean time and $y$-direction, respectively.
To leading order in $1/N$ this correlation function is given by simple
iteration of the bubble graph. After some lengthy but straightforward
calculation we obtain, to order ${\bf p}^2$,
\begin{equation}
D_\sigma(\omega=0,{\bf p})={\pi\over N}
\Bigl[{\overline{C}_2(\theta)L_c(\theta)\over 2}{\bf p}^2
+aM\tau\Bigr]^{-1}\, ,
\label{sprop}
\end{equation}
where $a=1$ when $\tau>0$ and $a=-1/2$ when $\tau<0$. At the
criticality, or $\tau=0$, $D_\sigma(\omega=0,{\bf p})\propto
{\bf p}^{-2}$, which implies $\eta=0$. The susceptibility $\chi$
is proportional to $D_\sigma(\omega=0,{\bf p}=0)\propto \tau^{-1}$
in both phases. Thus, we have $\gamma=\gamma'=1$. Eq.(\ref{sprop})
also implies that the correlation length $\xi$ has the form
\begin{equation}
\xi=\sqrt{\overline{C}_2(\theta)L_c(\theta)\over 2a M}\,
|\tau|^{-1/2}\, ,
\end{equation}
or equivalently $\nu=\nu'=1/2$.

  In order to obtain  the last exponent $\delta$ we need to introduce
an uniform external field $h$ coupled linearly to $\bar{\psi}\psi$
(or a mass term). In the presence of $h$ the gap equation to leading
order in $1/N$ is still given by Eq.(\ref{dveff3}) except for the
replacement $\sigma\rightarrow\sigma+h$ in the one-loop part. As a
consequence of this replacement the gap equation now becomes
\begin{equation}
m\biggl\{\tau+{1\over ML_c(\theta)}
\sum_{\nu=1}^\infty C_{2\nu}(\theta)
\Bigl(m(L,\theta)L\Bigr)^{2\nu}\biggr\}=h\, .
\end{equation}
At the critical point we have $m^3\propto h$, or in other words,
$\delta=3$.

  This completes the calculation of the exponents in 2+1 dimensions.
The same result was obtained in Ref. \cite{rosenstein} for $\theta=0$.
One obtains identical exponents by a similar calculation in
1+1 dimensions. We summarize the final result in Table \ref{tab1}.
It hardly escapes notice that the exponents in Table \ref{tab1}
are of the mean-field type. This does not constitute a surprise
 since we only work up to the leading order in $1/N$.
However, one should be aware that the large-$N$ limit alone does
not guarantee a mean-field result. For example, the non-linear
$\sigma$-model in lower than 4 dimensions violates mean-field
universality even in leading order in $1/N$ \cite{itzykson}.
The real reason in our case is that no $\sigma$-loop contribution
is involved to leading order in $1/N$. Of course, we anticipate
that the degeneracy of the critical exponents in 1+1 and 2+1
dimensions is lifted when we include higher order corrections.
It perhaps should be mentioned explicitly that, though the critical
exponents are independent of $\theta$, the amplitudes do depend on
$\theta$.

  Finally, we would like to point out that, if we regard
$\theta$ as the independent variable and fix $1/L$, we find that
the functional dependence on $\theta$ of some physical observables, such
as the mass gap and the induced current, near the critical region
depends on the value of $1/L$, as explicitly indicated in
Fig.\ref{fig2} and Fig.\ref{fig4}, with $L=L_c(0)$ as
a special point.

\section{The Nature of the quasi-particle}

  Since the interactions between quasi-particles are of the order
$1/N$, which is vanishingly small in the large $N$ limit, we are
allowed to discard the interaction all together when we attempt  to
elucidate the nature of the quasi-particle. The effective
Lagrangian is simply given by
\begin{equation}
{\cal L}_{\rm eff}=
\bar{\psi}[i\partial_\mu-a_\mu(\theta)]\gamma^\mu\psi
-m\bar{\psi}\psi \, ,
\end{equation}
with $m=m(L,\theta)$ and $\psi(x+L)=-\psi(x)$. The constant
gauge field can be eliminated by a gauge transformation, as mentioned
in the Introduction, at the expense of the $\psi$-field becoming
multi-valued when $\theta\neq 0$ or $1/2$. This multi-valuedness
signals that the quasi-particle is anyon-like.

  To be more specific, let us calculate the propagator for the
$\psi$-field, which can be easily written down in momentum space as
\begin{equation}
\tilde{S}(k_n,\omega;\theta)=i
{\omega\gamma_0-\tilde{k}_n\gamma_1+m \over
\omega^2-\tilde{k}_n^2-m^2+i\epsilon} \, .
\end{equation}
The propagator in coordinate space is then obtained by a
Fourier transform
\begin{equation}
S(x,t;\theta)={1\over L}\sum_{n=-\infty}^\infty
\int_{-\infty}^\infty {d\omega\over 2\pi} \,
e^{i(\omega t-k_n x)}\tilde{S}(k_n,\omega;\theta)
=\bigl[(i\partial_\mu-a_\mu(\theta))\gamma^\mu+m\bigr]
S_0(x,t;\theta) \, ,
\label{prop}
\end{equation}
where  $S_0(x,t;\theta)$ is the ``bosonic'' propagator
\begin{eqnarray}
S_0(x,t;\theta)&=&{1\over L}\sum_{n=-\infty}^\infty e^{-i k_n x}
\int_{-\infty}^\infty {d\omega\over 2\pi} e^{i\omega t}
{i\over \omega^2-\tilde{k}_n^2-m^2+i\epsilon} \, , \\
&=&{1\over L}\sum_{n=-\infty}^\infty \,
e^{-i k_n x}{1\over 2E_n}\left[\theta(t) e^{-i E_n t}
+\theta(-t) e^{i E_n t} \right] \, ,
\end{eqnarray}
and $E_n=\sqrt{\tilde{k}_n^2+m^2}$.
To carry out the discrete momentum sum we use the Poisson
summation formula
\begin{equation}
\sum_{n=-\infty}^\infty\, f(n)=\sum_{l=-\infty}^\infty
\int_{-\infty}^\infty d\phi \, e^{-i2\pi l\phi} f(\phi) \, .
\end{equation}
It is then straightforward, following the procedure in
\cite{bogoliubov}, to obtain
\begin{equation}
S_0(x,t;\theta)=\sum_{l=-\infty}^\infty
e^{i\pi l}\, e^{i2\pi(l+x/L)\theta} S_B(x+lL,t) \, ,
\label{prop0}
\end{equation}
where $S_B(x,t)$ is given by
\begin{equation}
S_B(x,t)={1\over 2\pi}\theta(x^2-t^2)K_0(m\sqrt{x^2-t^2})
-i{1\over 4}\theta(t^2-x^2)H_0^{(2)}(m\sqrt{t^2-x^2}) \, .
\label{propb}
\end{equation}
In the above equation $t$ should be understood as
$t(1-i\epsilon)\equiv t-i\epsilon\text{sign}(t)$ or equivalently
$t^2\rightarrow t^2-i\epsilon$,
to avoid divergences on the light-cone.

  The physical meaning of each term in $S_0(x,t;\theta)$
can be easily recognized. The ``$l$'' sum is obviously a sum
over winding numbers. The first factor is due to the anti-periodic
boundary condition. The second factor accounts for the total
flux embraced by a quasi-particle with a path which winds around
the origin ``$l$'' times. The third factor is nothing but the
relativistic boson propagator in flat 1+1 dimensional Minkowski
space from point $x_i=(0,0)$ to point $x_f=(x+lL,t)$. Therefore,
the above equation is a decomposition of $S_0(x,t;\theta)$ into
a sum  over paths of distinct homotopy class in the universal
covering space with $\theta$ taking the role of the statistics parameter. In
the limit $L>>t$ all terms except $l=0$ are exponentially (or
algebraically when $m=0$) suppressed and $S(x,t;\theta)$ approaches
the usual boson propagator.

  The interpretation of $\theta$ as  the statistics parameter
can be also understood from the point of view of
interchanging a pair of quasi-particles.
Since these quasi-particles can not penetrate through each other
due to the hard core (inherent from the fermionic Lagrangian when
$\theta=0$), the only way to interchange the positions of
quasi-particle number 1 located at $x_1=0$ and quasi-particle number
2 at $x_2=x$ is to let quasi-particle 1 travel through the interval
$(0,x)$, while quasi-particle 2 travels through the interval $(x,L)$
and then $(L:=0,x)$. The combined world line exactly circumscribes  the
ring once and therefore accumulates a phase factor
\begin{equation}
\exp\biggl\{i\oint dx^\mu a_\mu(\theta)\biggr\}=e^{i2\pi\theta} \, .
\end{equation}

In possession of the explicit expression for $S(x,t;\theta)$ we can use
the Bjorken-Johnson-Low formula \cite{bjl} to calculate the
equal-time anti-commutation relation
\begin{equation}
\{\psi_\alpha(x,0),\psi_\beta^\dagger(0,0)\}=
\Bigl(\lim_{t\rightarrow 0^+} S(x,t;\theta)\gamma_0-
\lim_{t\rightarrow 0^-} S(x,t;\theta)\gamma_0\Bigr)_{\alpha\beta}
\end{equation}
and see whether  the usual anti-commutator is affected by the presence
of the magnetic flux. It is readily verified that
\begin{equation}
\lim_{\epsilon\rightarrow 0}\lim_{t\rightarrow 0}
S(x,t(1-i\epsilon);\theta)={1\over 2}{\rm sign}(t)
\sum_{l=-\infty}^\infty (-1)^l \delta(x+lL)\gamma_0 + \,\,
\text{terms independent of $t$} \, ,
\end{equation}
which in turn yields
\begin{equation}
\{\psi_\alpha(x,0),\psi_\beta^\dagger(0,0)\}=
\delta_{\alpha\beta}\sum_{l=-\infty}^\infty (-1)^l \delta(x+lL) \, .
\end{equation}
While the sum over ``$l$'' is again due to the geometry of the ring,
we have verified that $\theta$ does not enter the canonical equal-time
anti-commutator in our model, in contrast to the graded commutator
obeyed by anyons induced via the Chern-Simons action. This result
should not be surprising  in light of the local character of the
canonical commutation relation and the fact that the magnetic flux
is global and never attached to each particle in our model. On the
other hand, the propagator contains global information and should
have a non-trivial $\theta$-dependence, which in turn yields the
non-trivial dependence on $\theta$ of  thermodynamical quantities.
Thus, the nature of the quasi-particle now becomes clear.
Locally, the quasi-particle is a fermion, but globally, the
quasi-particle behaves like an anyon.

  Similarly, the propagator in 2+1 dimensions can be calculated and
is again given by Eq.(\ref{prop}) except that $S_0(x,t;\theta)$ in
Eq.(\ref{prop0}) should be replaced by the 2+1 dimensional version
\begin{equation}
S_0(x,y,t;\theta)=\sum_{l=-\infty}^\infty
e^{i\pi l}\, e^{i2\pi(l+x/L)\theta} S_B(x+lL,y,t) \, ,
\end{equation}
with
\begin{equation}
S_B(x,y,t;\theta)=
\theta(r^2-t^2){e^{-m\sqrt{r^2-t^2}}\over 4\pi\sqrt{r^2-t^2}}
-i\theta(t^2-r^2){e^{-im\sqrt{t^2-r^2}}\over 4\pi\sqrt{t^2-r^2}} \, ,
\end{equation}
and $r=\sqrt{x^2+y^2}$.
Using the identity
\begin{equation}
\lim_{\epsilon\rightarrow 0}{\epsilon\over(r^2+\epsilon^2)^{3/2}}
=2\pi\delta(x)\delta(y) \, ,
\end{equation}
the anti-commutator in 2+1 dimensions is then given by
\begin{equation}
\{\psi_\alpha(x,y,t),\psi_\beta^\dagger(0,0,t)\}=\delta_{\alpha\beta}
\sum_{l=-\infty}^\infty (-1)^l \delta(x+lL)\delta(y) \, .
\end{equation}
It is not difficult to see that the quasi-particle in the 2+1 dimensional
case is anyon-like only in the $x$ direction. In other words, only
those world lines which circumscribe the axis of the cylinder
non-trivially can acquire phases of integer multiples of
$2\pi\theta$.

\section{Summary and outlook}

  We considered the statistical mechanics of the Gross-Neveu
model on a ring and on a cylinder with a magnetic solenoid
coinciding with the axis. Among the interesting results we obtained
are 1) the periodic but non-analytic dependence of thermodynamical
observables on the magnetic flux ($\theta$) and
2) the existence of an interval
of $\theta\in(1/3,2/3)$ (modulo integers) where parity is
always spontaneously broken. All these phenomena are  explicit
manifestations of the Aharonov-Bohm effect in thermodynamics or in
many-body systems. We further showed that the mean-field nature of
the existing phase transitions is preserved to leading order in
$1/N$, by verifying the $\theta$-independence of all the critical
exponents. The precise nature of the quasi-particle, locally
fermion-like and globally anyon-like, was illuminated through the
calculation of the equal-time commutator and the decomposition of
the propagator into a sum  over paths of distinct homotopy class
or winding number.

  There are several directions in which  our present work
can be extended within
the same model. The most interesting one is to include an external
space-time dependent electromagnetic field and study the response
to it. To set up junctions and investigate tunnelings and
interferences could also be interesting.
Although we only worked to  leading order in $1/N$, we
do expect that our results, especially in the 2+1 dimensional case,
are qualitatively stable against higher order $1/N$-corrections,
based on the next to leading order study of the same model in Ref.
\cite{gat} in the absence of $\theta$. Nevertheless, an explicit
calculation of the higher order corrections should be carried out.
One may also try to introduce a chemical potential and study the
situation away from the half-filling point.

  As we have been emphasizing all along, the periodic but
non-analytic $\theta$-dependence of thermodynamical observables
is a manifestation of the Aharonov-Bohm effect in many-body systems
with phase transitions, not necessarily pertinent only to the
Gross-Neveu model itself. For this reason we anticipate that similar
phenomena can be found in other models. The detailed microscopic
interactions are likely to play minor role, as long as phase
transitions exist. Thus, in order to make direct contact with
experimental verification, it would be very interesting to consider
more realistic models.

\begin{figure}
\caption{Mass gap $m(L,\theta)$ (in units of $M$) as a
function of $1/L$ for $\theta=0$ (fermion limit), 0.3, 0.4
and 1/2 (boson limit) in the 1+1 dimensional Gross-Neveu model.
Also plotted are the first order (dotted line) and second order
(dashed line) perturbative results. The sudden turns of the dotted
and dashed lines for small values of $1/L$ reflect the fact that
the perturbative running coupling diverges at $L=4L_c(0)$.}
\label{fig1}
\end{figure}

\begin{figure}
\caption{Mass gap $m(L,\theta)$ as a function of $\theta$ for
$L=1$, $L=L_c(0)=\ln{4}$ and $L=2$ (in units of $M$). The functional
form of the $\theta$-dependence of $m(L,\theta)$ in relevant
regions is also indicated explicitly.}
\label{fig2}
\end{figure}

\begin{figure}
\caption{Phase diagram in $(1/L,\theta)$-plane (with $M=1$).
In the interval of $\theta\in (1/3,2/3)$ the system is always
in the parity broken phase, independent of the value of $1/L$.}
\label{fig3}
\end{figure}

\begin{figure}
\caption{Induced current $J[L,\theta]$ as a function of $\theta$
for $L=1$, $L=L_c(0)=\ln{4}$ and $L=2$ (with $M=1$). The functional
form of the $\theta$-dependence of $J[L,\theta]$ in relevant
regions is also indicated explicitly.}
\label{fig4}
\end{figure}

\begin{table}
\caption{critical exponents}
\begin{tabular}{cccccccccc}
    & $\alpha$ & $\alpha'$ & $\beta$ & $\gamma$ & $\gamma'$ &
      $\delta$ & $\nu$ & $\nu'$ & $\eta$ \\
\tableline
D=1+1, $\theta\in [0,1/2)$ or $(1/2,1]$
&0     &  0    &1/2  &  1   &   1   &3&1/2&1/2& 0 \\
D=2+1, $\theta\in [0,1/3)$ or $(2/3,1]$
&0     &  0    &1/2  &  1   &   1   &3&1/2&1/2& 0
\label{tab1}
\end{tabular}
\end{table}


\begin{references}

\bibitem{shapere} R.~E.~Prange and S.~M.~Girvin,  The
  Quantum Hall Effect  (Springer, New York, 1987);

  A.~Shapere and F.~Wilczek, Geometric Phases
   in Physics (World Scientific, Singapore, 1989)

\bibitem{wilczek} F.~Wilczek, Fractional Statistics and Anyon
   Superconductivity (World Scientific, Singapore, 1990)

\bibitem{aharonov} Y.~Aharonov and D.~Bohm, Phys. Rev. 115
      (1959) 485

\bibitem{gross} D.~J.~Gross and A.~Neveu, Phys. Rev. D10 (1974) 3235

\bibitem{hubbard} E.~Fradkin, Field Theories of Condensed
  Matter Systems (Addison-Wesley, Redwood City, 1991) chapter 4

\bibitem{wen} X.~G.~Wen, F.~Wilczek and A.~Zee, Phys. Rev.
   B39 (1989) 11413

\bibitem{luescher} M.~L\"uscher, Phys. Letters B118 (1982) 391

\bibitem{luescher1} M.~L\"uscher, Nucl. Phys. B219 (1983) 233

\bibitem{baal}
   P.~van~Baal and A.~S.~Kronfeld, Nucl. Phys. B (Proc. Suppl.) 9
          (1989) 227;

   H.~Hansson, P.~van~Baal and I.~Zahed, Nucl. Phys. B289 (1987) 628;

   J.~Koller and P.~van~Baal, Nucl. Phys. B302 (1988) 1

\bibitem{thooft} G.~'t Hooft, Nucl. Phys. B153 (1979) 141

\bibitem{twisted} A.~Coste, A.~Gonzalez-Arroyo, C.~P.~Korthals-Altes,
    B.~S\"oderberg and A.~Tarancon, Nucl. Phys. B287 (1987) 569;

    M.~Garcia Perez, A.~Gonzalez-Arroyo and P.~Martinez,
    preprints INLO-PUB-18/93, FTUAM-93/40 (1993)

\bibitem{dashen} R.~F.~Dashen, S.~K.~Ma and R.~Rajaraman,
   Phys. Rev. D11 (1975) 1499

\bibitem{review} B.~Rosenstein, B.~J.~Warr and S.~H.~Park,
   Phys. Rep. 205 (1991) 59

\bibitem{leinaas} J.~M.~Leinaas and J.~Myrheim, Il Nuovo Cimento
   37B (1977) 1

\bibitem{gerry} C.~C.~Gerry and V.~A.~Singh, Il Nuovo Cimento
   73B (1983) 161

\bibitem{arovas} D.~Arovas, J.~R.~Schrieffer, F.~Wilczek and
   A.~Zee, Nucl. Phys. B251 (1985) 117

\bibitem{itzykson} C.~Itzykson and J.-M.~Drouffe,  Statistical
   Field Theory (Cambridge Univ. Press, Cambridge, 1989)

\bibitem{rosenstein} B.~Rosenstein, B.~J.~Warr and S.~H.~Park,
   Phys. Rev. D39 (1989) 3088

\bibitem{bogoliubov} N.~N.~Bogoliubov and D.~V.~Shirkov,
  An Introduction to Theory of Quantized Fields
   (Wiley-Interscience, New York, 1959)

\bibitem{bjl} J.~D.~Bjorken, Phys. Rev. 148 (1966) 1467;

   K.~Johnson and F.~E.~Low, Prog. Theor. Phys. (Kyoto) Supp.
   37-38 (1966) 74

\bibitem{gat} G.~Gat, A.~Kovner, B.Rosenstein and B.~J.~Warr,
    Phys. Lett. B240 (1990) 158

\end{references}
\end{document}